\documentclass[preprintnumbers,prd,twocolumn,showpacs,floatfix,
preprintnumbers,letterpaper,amsmath,amssymb,nofootinbib,
superscriptaddress]{revtex4}
\usepackage{amsmath}
\usepackage{amsfonts}
\usepackage{amssymb}
\usepackage{latexsym}
\usepackage{graphicx}
\usepackage[english]{babel}

\usepackage{color}

\def\be{\begin{equation}}
\def\ee{\end{equation}}
\def\bea{\begin{eqnarray}}
\def\eea{\end{eqnarray}}

\begin{document}

\title{Entropic Force Scenarios and Eternal Inflation}

\author{Taotao Qiu}
 \email{xsjqiu@gmail.com}
 \affiliation{$1$. Department of Physics, Chung-Yuan Christian
University, Chung-li 320, Taiwan}
 \affiliation{$2$. Department of Physics and Center for Theoretical
Sciences, National Taiwan University, Taipei 10617, Taiwan}
\affiliation{$3$. Leung Center for Cosmology and Particle
Astrophysics National Taiwan University, Taipei 106, Taiwan}

\author{Emmanuel N. Saridakis}
\email{Emmanuel_Saridakis@baylor.edu}
 \affiliation{CASPER,
Physics Department, Baylor University, Waco, TX  76798-7310, USA}
\affiliation{National Center for Theoretical Sciences, Hsinchu,
Taiwan 300 }

\begin{abstract}
We examine various entropic inflation scenarios, under the light of
eternality. After describing the inflation realization and the
normal condition for inflation to last at the background level, we
investigate the conditions for eternal inflation with the effect of
thermal fluctuations produced from standard radiation and from the
holographic screen. Furthermore, we incorporate stochastic quantum
fluctuations through a phenomenological, Langevin analysis, studying
whether they can affect the inflation eternality. In
one-holographic-screen scenarios eternality can be easily obtained,
while in double-screen considerations inflation is eternal only in
the high-energy regime. Thus, from the cosmological point of view,
one should take these into account before he can consider entropic
gravity as a candidate for the description of nature. However, form
the string theory point of view, inflation eternality may form the
background for the ``Landscape'' of string/M theory vacua, leading
to new perspectives in entropy gravity.

\end{abstract}

\pacs{98.80.-k, 98.80.Cq, 04.50.-h}

\maketitle

\section{Introduction}

After almost three decades of extensive research, inflation is now
considered to be a crucial part of the cosmological history of the
universe \cite{inflation}, having affected indelibly its
observational features. The inflation paradigm, in conventional as
well as in higher dimensional frameworks
\cite{Lindebook,LidseyRMP,Baumann:2009ni}, can successfully solve
theoretical problems such as the flatness, the horizon and the
monopole ones, and moreover it provides the right behavior of
primordial fluctuations with a nearly scale-invariant power
spectrum.

One important issue that has to be encountered in inflationary
cosmology is that of eternality, that is whether inflation has a
beginning and/or an end \cite{Vilenkin:1983xq,Guth:2007ng}. In
general there are two different points of view for its
incorporation. From the cosmological side a picture that there are
always parts of the universe that inflate and parts that have exit
inflation, like our own observable universe part, in a procedure
without beginning, offers a way out from the singularity
\cite{Hawking:1973uf,Borde:1993xh} and transplanckian
\cite{Martin:2000xs} problems. However, in order for such a
procedure to be the case in Nature, one still needs to find a
mechanism that can lead areas of the universe of non-zero measure to
exit inflation, since in our ``local'' part inflation has obviously
ended. We stress that these parts must be of non-zero measure since
a minimal requirement that inflation has to end at least in one part
amongst infinite eternally inflating areas, leads to an intense
fine-tuned and anthropic description of Nature \cite{Guth:2000ka}.
On the other hand, from the String Theory point of view, eternal
inflation may form the background for (or be formed by) the
``Landscape'' of string/M theory vacua \cite{Kachru:2003aw}, induced
by flux compactification \cite{Bousso:2000xa} or other mechanisms
\cite{Susskind:2003kw,Douglas:2003um}, that is for the existence of
a huge number of possible (false) vacua. Thus, from this point of
view inflation eternality has an additional reason of being a
desirable feature of a cosmological theory, and furthermore one does
not need to worry so much about the inflation exit, since it is
adequate for it to be realized in only one area of the universe
(equivalent of choosing a specific vacuum along a landscape of $\sim
10^{500}$ ones \cite{Douglas:2003um}).

Recently, an extended holographic picture was suggested by Verlinde
\cite{Verlinde:2010hp}, in which gravity is no longer a fundamental
theory, but emerges from a statistic effect of a holographic screen
(a similar scenario was discussed by Padmanabhan in
\cite{Padmanabhan:2009vy}, based on the earlier considerations of
\cite{Padmanabhan:2004qc}). Such an ``entropic'' origin of gravity
was based on the holographic principle, conjectured as a significant
property of quantum gravity, stating that physics of a volume of
space is encoded on its boundary, such as the gravitational horizon
\cite{'tHooft:1993gx}. Although there is a controversy in the
foundations of entropic gravity itself \cite{Gao:2010yy}, the idea
is very interesting and the cosmological implications of its various
scenarios were extensively studied in the literature. Besides the
original formulation (see \cite{Cai:2010hk} and references therein),
people proposed various inflation models of which inflation was
assumed to be driven by one or two holographic screen(s). In
one-screen inflation model, the holographic screen acts as a
boundary term in the Einstein equation, which can force the universe
to accelerated expansion even if the universe is radiation-dominated
\cite{Easson:2010xf}. In two-screen model, an ``inner" Schwarzschild
screen was added inside the holographic screen, which made the
inflation realization easier \cite{Cai:2010zw,Cai:2010kp}.
Additionally, there are many other implications of entropic gravity,
such as to explain late-time acceleration
\cite{Easson:2010av,Gao:2010fw} and  to incorporate black holes
\cite{Myung:2010jv}.

In the present work we focus on the eternality issue of inflation
realization in the various entropic scenarios of the literature,
both at background and fluctuation levels. In particular, we examine
on what conditions inflation can be eternal, and whether its
eternality can be prevented by stochastic quantum fluctuations. The
plan of the work is as follows. In section \ref{Verl} we briefly
review Verlinde's basic entropic gravity formulation. In section
\ref{1screenm} we examine the eternality for single-screen inflation
scenario, while in section \ref{2screenm} we study the case of
double-screen inflation model. Finally, section \ref{Conclusions} is
devoted to the summary of the results.

Lastly, we mention that in the whole manuscript  in order to
simplify the symbolism  we set the light speed $c$ and the reduced
Planck $\hbar$ and Boltzmann $k_B$ constants to 1, but we keep the
Newton's constant explicitly for clarity.

\section{Verlinde's basic scenario}
\label{Verl}

The basic ingredient of Verlinde's idea is that the boundary physics
can be described by thermodynamics satisfying a holographic
distribution, where the number of degrees of freedom on this
holographic screen is proportional to its area, that is \be
 dN=\frac{dA}{G},
 \ee
 where $G$ is the Newtonian gravitational
constant. Thus, the classical holographic entropy on this screen is
given by \be
 \label{Sb}
 S_b=\frac{A}{4G}=\frac{ \pi r_b^2}{  G} ~,
\ee where $r_b$ is the radius location of the boundary surface
${\cal S}$. Variation of energy with respect to the radius will
provide the entropic force \cite{Verlinde:2010hp}:
 \be
 \label{Fe}
 F_{e}=-\left(\frac{dE}{dr}\right)_b=-\left(T\frac{dS}{dr}\right)_b
=-2\pi \frac{T_br_b}{ G}~,
 \ee in which $T_b$ is the temperature of the
boundary of the system. Finally, due to the Unruh effect (when a
test particle with mass $m$ is located nearby the holographic screen
the variation of the entropy on this screen with respect to the
radius takes the form $\frac{dS}{dr}=-2\pi  m$) the above force
yields an entropic acceleration $a_e$ of the form
\cite{Unruh:1976db}:
 \be
\label{ae}
 a_e\equiv
\frac{F_e}{m}=2\pi T_b~. \ee Note that the corresponding entropic
pressure is negative $P_e=\frac{F_e}{A_b}=-\frac{1}{2
G}\frac{T_b}{r_b}$, and so it is expected to realize an accelerating
process of the universe.

The above simple analysis can be straightforwardly generalized to
the full relativistic case. The natural generalization of Newton's
potential is $ \phi ={1\over 2} \log(\xi^a\xi_a) $, where $\xi^a$ is
a global time-like Killing vector  \footnote{Note that in the
present manuscript we follow the $(+,-,-,-)$ metric signatures, and
thus in the above definition we have inverted the sign inside the
logarithm comparing to the original expression in
\cite{Verlinde:2010hp}.}, and the exponential $e^\phi$ represents
the redshift factor that relates the local time coordinate to that
at a reference point with $\phi=0$, which we will take to be at
infinity. One considers a holographic screen on a closed surface of
constant redshift $\phi$, enclosing a certain static mass
configuration with total mass $M$ \cite{Verlinde:2010hp}. Due to
equipartition $ M={1\over 2}\int_{\cal S}TdN$, where the Unruh
temperature now writes as \be \label{temperature}
T=\frac{1}{2\pi}e^\phi N\cdot\nabla\phi. \ee
 Therefore, for the total mass we obtain $4\pi
GM = \int_{\cal S}  e^{\phi} \nabla \phi\cdot dA $, or expressed in
terms of the Killing vector: \be M ={1\over 8\pi G}\int_{\cal S}
dx^a \!\wedge \!dx^b\, \epsilon_{abcd} \nabla^{c}\xi^{d}~. \ee Thus,
expressing the total mass in terms of the energy-momentum tensor we
result to the Einstein equations, namely \be \label{Einstein}
2\int_{\Sigma}\left(T_{ab} -{1\over 2}Tg_{ab}\right) n^a \xi^b dV=
{1\over 4\pi G}\int_{\Sigma}R_{ab} n^a \xi^b dV~, \ee where $\Sigma$
is the three dimensional volume bounded by the holographic screen
$\cal S$ and $n^a$ is its normal. Since this relation can hold for
all appropriate Killing vectors and for arbitrary screens
\cite{Verlinde:2010hp}, it is sufficient in providing the full
Einstein equations.

The above formulation of gravity as an entropic force, which could
therefore lead to descriptions of the inflationary as well as the
dark energy epoches, is quite general. Nonetheless, one still needs
to construct more precise and detailed models in order to proceed to
a quantitative analysis of various cosmological eras. Concerning
inflation, which is the subject of the present work, we stress here
that the entropic origin of gravity itself is consistent with both
finite or eternal behavior, similar to the case of standard
gravitational theories. Thus, the eternality subject has to be
examined in each explicit inflationary model separately, since it
depends on the details of each scenario. This is performed in the
next two sections.

\section{The single-screen Model}
\label{1screenm}

Although the qualitative features of inflation in the entropic
context were discussed in \cite{Wang:2010jmb}, the first explicit
quantitative inflation model was proposed in \cite{Easson:2010xf}.
The basic ingredient of this model is that acceleration is driven by
an additional surface term in the Einstein's equations, which comes
from the Holographic screen assumed by Verlinde. In this section,
after briefly reviewing the model, we firstly examine the condition
for inflation to last classically, then we study its eternality
incorporating thermal fluctuations, and finally we perform a
detailed Langevin analysis, as a first approach to the stochastic
quantum effects. In the following we assume a flat
Friedmann-Robertson-Walker (FRW) background geometry with metric \be
ds^2=  dt^2-a^2(t)\,\delta_{ij} dx^i dx^j~, \ee where $t$ is the
cosmic time, $x^i$ are the comoving spatial coordinates, and $a(t)$
is the scale factor. We also introduce the Hubble parameter
$H=\dot{a}/a$, with a dot denoting differentiation with respect to
$t$.

In single-screen inflation model, one of the two Friedmann equations
reads
 \be
\frac{\ddot a}{a}=-\frac{4\pi
 G}{3}\left(\rho+3P\right)+C_HH^2+C_{\dot H}\dot H~, \ee where
$C_H$ and $C_{\dot H}$ are dimensionless coefficients, which are
expected to be bounded by $C_H<1$ and $0\leq C_{\dot H}\leq 3/4\pi$
according to the authors who first constructed the
scenario \cite{Easson:2010xf}\footnote{We mention here that the same
authors, when they apply the same scenario in the late-time accelerated
epoch instead of inflation, they use a different lower limit for $C_H$
\cite{Easson:2010av,Casadio:2010fs}. Although this behavior would require
an explanation by those authors, for the purpose of the present work the
lower limit of $C_H$ is irrelevant, since only the sign of $C_H-1$ plays a
role, that is its upper limit, which is always $1$ in all works.}.
Moreover, taking into account higher order
corrections to the entropy, the above relation becomes:
\be\label{friedmann1}
 \frac{\ddot a}{a}=-\frac{4\pi
G}{3}\left(\rho+3P\right)+C_HH^2+C_{\dot H}\dot H+g\frac{  G}{\pi
}H^4~, \ee
 where the dimensionless factor $g$ incorporates the
effective number of independent degrees of freedom which is
dimensionless (note that since we have set  $c=\hbar=k_B=1$, $GH^2$ is
dimensionless and thus all the formulas have are homogeneous in units).
 On the other hand,
the second Friedmann equation writes as
 \be
\label{friedmann2}
 H^2=\frac{8\pi G}{3}\rho+C_HH^2+C_{\dot H}\dot H~. \ee Amazingly
enough, the combination of (\ref{friedmann1}) and (\ref{friedmann2})
provides us the modified second Friedmann equation independent on
$C_H$ and $C_{\dot H}$, namely
 \be \dot H=-4\pi
G \left(\rho+P\right)+g\frac{ G}{\pi }H^4~. \ee
  Moreover, differentiating
(\ref{friedmann2}) with respect to $t$ and ignoring the higher order
derivative of $H$ under slow-roll approximation, we have the
continuity equation:
 \be
\label{continuity1} \dot\rho+3H(1-C_H)\left(\rho+P\right)=0~, \ee
(note that this relation can be viewed as originating from the
well-known generalized conservation equation in which matter
exchanges energy with vacuum).
 Finally, the expression for the
entropic pressure that drives inflation is:
 \be
P_e=
-\frac{2}{3}\rho_{c0}\left(\frac{H^2}{H^2_0}+\frac{gG}{\pi}\frac{H^4}{
H^2_0}\right)~,\ee where we have introduced the critical energy
density $\rho_c=3H^2/(8\pi G)$, with the subscript $0$ denoting the
present-day value of a quantity. Clearly, when $gH^4/\pi=4\pi
(\rho+P)$, which in the case of relativistically high energies
becomes $gH^4/\pi=16\pi\rho/3$, we acquire $\dot H=0$ and thus a
de-Sitter expanding phase can be obtained \cite{Easson:2010xf}.

Before proceeding to the investigation of the eternality issues of
the model at hand, let us make some comments. Firstly, note that the
last two terms in Friedmann equation (\ref{friedmann2}) can be
viewed as an ``effective" inflation part, which can drive
acceleration. Straightforwardly, the same  equation can also be
applied to the dark-energy epoch, that is to explain the late-time
acceleration, as it is done in the so-called ``running cosmological
constant'' scenarios \cite{Sola:2007sv,Basilakos:2009wi} (though
from a different ideological point of view). In this approach the
time-dependent part of the effective cosmological constant
$\Lambda(t)$ can be made to be $\propto H^2$, which is similar to
the case at hand except for the $\dot H$ term. However, and more
importantly, $\Lambda(t)$ should also contain a non-vanishing
constant additive term, in order to fit the combined observational
data \cite{Basilakos:2009wi}. This is a disadvantage of the present
model, since one should provide an explanation for the modification
of Friedmann equation (\ref{friedmann2}) between early and late
times, in order to acquire a realistic model\footnote{We thank an
anonymous referee for pointing this out.}. In the same lines, as it was
mentioned above, one
could put into question the   bounds $C_H<1$ and $0\leq
C_{\dot H}\leq 3/4\pi$ \cite{Easson:2010xf} that the authors give
for their model. In summary, the one-screen
model examined here seems to have ambiguities concerning the correct
quantitative behavior. Clearly one could study a generalization,
including a constant term in (\ref{friedmann2}), or abandon the
aforementioned bounds in the model parameters. However, in the
present work we prefer to remain in the original version of the
scenario, and examine it under the eternality point of view, instead
of trying to improve it first, which could be the subject of
interest of a separate work.

\subsection{Inflation Eternality: Background Analysis}

Let us now discuss the inflation eternality realization in
one-screen entropic scenario. Firstly we have to determine the
condition for a  {\it global} inflation exit, that is with no part
of the universe going on inflating. In the scalar-field driven
inflation models this is just the slow-roll condition. In the
scenario at hand there is no field slow-rolling, however we can
still use the definition of the ``slow-roll" parameter
$\epsilon\equiv-\dot H/H^2$,  and thus the condition for inflation
lasting remains $\epsilon\lesssim 1$ (this condition arises from the
definition of inflation and thus it is model-independent), or
equivalently
 \be
\label{slowroll} \dot H+H^2\gtrsim 0~. \ee

Starting from equations (\ref{friedmann1}) and (\ref{friedmann2})
and assuming that the universe is filled with radiation
($P=\rho/3$), we can eliminate $\rho$ and obtain an equation with
$H(t)$ only: \be
 \label{eq111} \dot
H(t)(1-2C_{\dot{H}})=2(C_H-1)H^2+g\frac{G}{\pi}H^4~. \ee
 Using this
equation, condition (\ref{slowroll}) imposes a constraint on $H(t)$,
namely: \be \label{condsr}
H\gtrsim\sqrt{\frac{\pi}{gG}\left[2C_{\dot{H}}-1+2(1-C_H)\right]}~.
\ee
 In summary, as long as $H(t)$ is larger than the value of the
right-hand-side of the above condition, inflation will not globally
exit, and thus we obtain its eternality realization at the
background level. Taking into account the allowed intervals of the
parameters, we deduce that this condition can be easily fulfilled.
However, one must also investigate the role of fluctuations
generated during inflation, on the eternality condition. This is
performed in the next subsections.

\subsection{Condition for Eternal Inflation: Thermal Fluctuation Analysis}

In the previous subsection we investigated the background evolution,
extracting the corresponding condition for inflation lasting in the
whole universe. Here we examine the role of thermal fluctuations.
Note that during inflation, the universe will expand to nearly
$e^3\approx 20$ causally independent Hubble-sized regions in one
Hubble time $\Delta t\approx H^{-1}$. In any of these regions the
energy density will be decreased by $\delta_c\rho=|\dot\rho|\Delta
t\approx|\dot\rho|H^{-1}$, however this may be corrected when we
take into account the fluctuations. In particular, if there is
non-zero probability that there exist regions of the universe where
the fluctuations are bigger than the corresponding classical change,
then in these regions inflation will be eternal, despite the fact
that in the other regions inflation will have end
\cite{Guth:2007ng}. In a holographic system, the fluctuations is
generated in a thermal way \cite{Easson:2010xf}, so we define the
condition for the inflation to be eternal as:
 \be\label{eternal1}
\delta_t\rho>\delta_c\rho, \ee
 where $\delta_t\rho$ is the thermal
fluctuation generated during inflation. We desire to emphasize that
this condition looks similar to the well-known condition for eternal
inflation driven by scalar field \cite{Guth:2007ng}:
$\delta_q\phi>\delta_c\phi$ \footnote{Note that the condition for
eternal inflation in case of non-commutativity or non-minimal
coupling has also been discussed, see \cite{Cai:2007et} and
\cite{Feng:2009kb} respectively.}, but the difference is that in
that case fluctuations comes from the quantum behavior of the scalar
field. From this comparison one can see that our condition is very
reasonable, too.

Using (\ref{continuity1}) and the Friedmann equations
(\ref{friedmann1}) and (\ref{friedmann2}), we can express the
classical change of the energy density $\delta_c\rho$ as \bea
\label{deltacrho1} \delta_c\rho=\frac{3(1-C_H)H^2}{2\pi
G(1-2C_{\dot{H}})}\left[ (1-C_H)-\frac
{gC_{\dot{H}}}{\pi}GH^2\right]~, \eea where we have also made use
that $\dot\rho=-4H\rho(1-C_H)$ (a relation that holds for
radiation-dominated universe) is always negative. On the other hand,
the thermal change $\delta_t\rho$ can be estimated through its
correlation function in position space, namely
$\delta_t\rho=\sqrt{\langle\delta\rho^2\rangle}$, with the relation
\cite{Magueijo:2002pg}:
 \be
\langle\delta\rho^2\rangle=C_V(R)\frac{T^2}{R^6}~, \label{deltarho}
\ee
 where
$C_V(R)\equiv\partial\langle E\rangle/\partial T$ is the heat
capacity in a sphere of radius $R$. Usually two kinds of thermal
fluctuations could be taken into account, namely thermal particle
fluctuations from normal radiation inside the bulk of the universe,
and holographic fluctuations from the boundary screen. According to
different fluctuating mechanisms, the condition for eternal
inflation will certainly be different. Thus, in the next two
paragraphs we will investigate both of them separately.

\subsubsection{Normal radiation}

The energy density of normal radiation inside the bulk of universe
can be written as \cite{Easson:2010xf} \be \rho_{tr}=4\sigma{\cal
G}(T)T^4, \ee where ${\cal G}(T)=45g$ is the effective number of
degrees of freedom at temperature $T$. The subscript ``tr'' stands
for ``thermal radiation''. Thus, from the heat capacity definition
we acquire \be C_V=960\pi g\sigma R_{tr}^3T^3~, \ee where $R_{tr}$
is the correlation length of radiation fluctuations,  given as usual
from $R_{tr}=c_s/H$, with $c_s$ the sound speed of the radiation.
Therefore, we get the following thermal fluctuation of the system:
$\delta_t\rho|_{_{tr}}=\sqrt{960\pi g\sigma T^5/R_{tr}^3}$. Setting
$T=H/2\pi$ as the Gibbons-Hawking temperature, we finally obtain \be
\label{deltaqrho11}
 \delta_t\rho|_{_{tr}}=\frac{1}{\pi^2}\sqrt{\frac{30
g\sigma}{c_s^3}}H^4~.
 \ee
From  (\ref{deltacrho1}) and (\ref{deltaqrho11}) we can see that in
single-screen model, the condition for eternal inflation
(\ref{eternal1}) with normal radiation fluctuation has the form:
 \be
\frac{1}{\pi^2}\sqrt{\frac{30
g\sigma}{c_s^3}}H^4>\frac{3(1-C_H)H^2}{2\pi G
(1-2C_{\dot{H}})}[(1-C_H)-\frac{gC_{\dot{H}}}{\pi}GH^2]~, \nonumber
\ee which gives the constraint on $H$: \be \label{cond1}
H>\frac{\sqrt{3\pi}(1-C_H)}{\sqrt{G}\left[2(1-2C_{\dot{H}})c_s^{-\frac{3}{2}}\sqrt{
30g\sigma}+3(1-C_H)gC_{\dot{H}}\right]^\frac{1}{2}}~. \ee

Finally, we mention here that (\ref{cond1}) has to be considered
along with (\ref{condsr}) in order for inflation to be eternal,
since when inflation globally exits, there will be no eternal
inflation. Thus, if the right-hand-side of (\ref{cond1}) is larger
than that of (\ref{condsr}), which depends on the parameter choice,
we can take (\ref{cond1}) as eternality condition safely, while if
it is not, (\ref{condsr}) should be taken as the eternal condition.
Again, we can see that the eternality conditions can be easily
fulfilled.

\subsubsection{Holographic screen}

Now we focus on the case that the thermal fluctuation is produced
holographically. The number of (finite) degrees of freedom on the
boundary screen is proportional to the surface area $\sim R_{hs}^2$,
where $R_{hs}$ is the correlation length and ``hs'' stands for
``holographic screen''. The energy of each degree of freedom is
roughly the temperature of the screen $T$ in thermal equibrilium,
thus the total energy of the holographic screen is $\langle
E\rangle\sim R_{hs}^2T/G$, and the heat capacity $C_V\sim
c_vR_{hs}^2/G$ where $c_v$ is a constant of the order of
${\cal{O}}(1)$ determined by the detailed microscopic quantities of
quantum gravity. Therefore, (\ref{deltarho}) gives
$\delta_t\rho|_{_{hs}}=T\sqrt{c_v/G}/R_{hs}^2$, and setting
$R_{hs}\simeq1/H$ and $T=H/2\pi$ \cite{Cai:2010kp} we finally obtain
\be \label{deltaqrho12}
\delta_t\rho|_{_{hs}}=\sqrt{\frac{c_v}{G}}\frac{H^3}{2\pi}~. \ee

From  (\ref{deltacrho1}) and (\ref{deltaqrho12}) we deduce that in
single-screen scenarios the condition for eternal inflation
(\ref{eternal1}), under holographic fluctuations, becomes:
 \be
\sqrt{\frac{c_v}{G}}\frac{H^3}{2\pi}>\frac{3(1-C_H)H^2}{2\pi G
(1-2C_{\dot{H}})}\left[(1-C_H)-\frac{C_{\dot{H}}g}{\pi}GH^2\right]~,
\nonumber \ee
 which
gives the constraint on $H$: \be\label{cond2} H>\frac{\pi
(1-2C_{\dot{H}})}{6gC_{\dot{H}}(1-C_H)\sqrt{G}}\left[\sqrt{c_v+\frac{36gC_{\dot{H}}
(1-C_H)^3}{\pi (1-2C_{\dot{H}})^2}}-\sqrt{c_v}\right],\ee
 and both
(\ref{cond2}) and (\ref{condsr}) have to be taken into account for
the same reason mentioned at the end of the above paragraph.

Before closing this subsection we have to make the following
comment. In the above two paragraphs we calculated separately the
thermal fluctuations from normal radiation and from the holographic
screen, and we gave the corresponding conditions for eternal
inflation, namely relations (\ref{cond1}) and (\ref{cond2})
respectively. However, although these are two independent
mechanisms, they can exist simultaneously and thus lead to a
combined effect, namely
$\delta_t\rho|_{_{tr}}+\delta_t\rho|_{_{hs}}>\delta_c\rho$. This
forces $H(t)$ to satisfy:
\begin{widetext}
 \be
\label{condtot} H>\frac{-\pi(1-2C_{\dot{H}})\sqrt{c_v}+\sqrt{\pi^2
c_v(1-2C_{\dot{H}})^2+12\pi(1-C_H)^2\left[2(1-2C_{\dot{H}})c_s^{-3/2}\sqrt{
30g\sigma} +3(1-C_H)gC_{\dot{H}}\right]}}{2\sqrt{G}
\left[2(1-2C_{\dot{H}})c_s^{-3/2}\sqrt{30g\sigma}
+3(1-C_H)gC_{\dot{H}}\right]}~\ee
\end{widetext} for eternal inflation.

\subsection{Effects from Quantum Corrections: Langevin Analysis}

In the previous subsection we investigated the conditions for
eternal inflation involving thermal fluctuations generated both from
normal radiation and holographic screen. Here we examine the
backreaction of the metric and the quantum fluctuations on the
background space-time, in order to see whether it can prevent
inflation from eternality. In particular, we proceed to an indirect
investigation of quantum fluctuations and incorporate them as a
stochastic effect, without caring about their specific microscopic
origin. Such an approach covers all possible effects of quantum
fluctuations at the phenomenological level, as is used in many
cosmological systems \cite{Feng:2009kb,Chen:2006hs}.

According to evolution equation (\ref{eq111}) and following
\cite{Chen:2006hs}, we formulate the overall cosmological evolution,
including the classical motion and the quantum fluctuations as a
stochastic effect modeled through a random walk, which can then be
described by a Langevin equation and analysis. In particular,
considering the system being perturbed by micro-fluctuations
described by a Gaussian white noise normalized as \be \label{mean}
\langle n(t)\rangle=0~,~~~\langle n(t)n(t')\rangle=\delta(t-t')~,\ee
we can write the Langevin equation for equation (\ref{eq111}) as:
 \be
\label{lagevin1}
 \dot H(t)=D_1H^2+D_2H^4+q_sn(t)~,\ee where we have
defined the coefficients
$D_1\equiv-\frac{2(1-C_H)}{1-2C_{\dot{H}}}$, $D_2\equiv\frac{gG}{\pi
(1-2C_{\dot{H}})}$, $q_s\equiv\frac{\varepsilon
G^{-3/4}}{1-2C_{\dot{H}}}$ (note that according to the bounds on
$C_H$, $C_{\dot{H}}$, we deduce that $D_1<0$ and $D_2>0$). Note also
that $D_1$ and $D_2$ have dimension of 0 and $-2$ respectively. In
the stochastic term the coefficient $G^{-3/4}$ is inserted as
usual for dimensional reasons, and $\varepsilon$ is a dimensionless
coefficient with rather small value \cite{Chen:2006hs}.

In equation (\ref{lagevin1}), if the last term on the right-hand
side is absent we recover the usual equation of motion (\ref{eq111})
and the system will follow a classical trajectory $H_c(t)$.
Therefore, we expand $H(t)$ around its classical value $H_c(t)$ up
to order $\mathcal{O}(q_s^2)$, namely
 \be
\label{expansion}
H(t)=H_c(t)+q_sG^{\frac{3}{4}}H_1(t)+q_s^2G^{\frac{3}{2}}H_2(t)+{\cal
O}(q_s^3)~,\ee where $G^{\frac{3}{4}}$ and $G^{\frac{3}{2}}$ have
been inserted for dimensional reasons. Substituting
this expansion into (\ref{lagevin1}) and setting the coefficients of
the $q_s$-powers to zero, we acquire the equations \bea \label{dhc1}
\dot H_c&=&D_1H_c^2+D_2H_c^4~,
\\ \dot H_1&=&2D_1H_1H_c+4D_2H_1H_c^3+G^{-\frac{3}{4}}n~,
\\
\dot H_2&=&D_1(H_1^2+2H_cH_2)+D_2(6H_c^2H_1^2+4H_c^3H_2).\ \
\label{dhc13} \eea

These equations cannot accept analytical solutions, however we can
obtain approximate solutions in the limit $|t/{\cal T}|\ll 1$ where
${\cal T}\gg H^{-1}$, that is performing our calculation within one
Hubble-interval ($t\sim H^{-1}$) \cite{Chen:2006hs}, with $t=0$ the time
where inflation starts, and then we impose the requirement the stochastic
fluctuations to be able to stop the eternal inflation. The explicit
calculations are presented in the Appendix \ref{appendix1}. We find that
stochastic fluctuations can hardly have any significant effect on
preventing inflation from eternality caused at the background or
thermal-fluctuation levels.

In summary, in this section we obtained the conditions for eternal
inflation in one-screen entropic inflation scenarios, taking into
account two types of thermal fluctuation productions. As we saw, for
general parameter choices the eternality conditions can be easily
fulfilled. We also performed a Langevin analysis and we found that
the quantum-stochastic effects from micro-fluctuations can hardly
change the behavior, and thus in the model at hand inflation has a
large probability to be eternal. Actually this result had already
been discussed in the original paper \cite{Easson:2010xf}, and thus
the authors resorted to quantum fluctuations in order to induce the
inflation exit, but the quantum-fluctuations incorporation remained
at the qualitative level. However, in this section we performed a
quantitative analysis and our conclusions seem trustworthy.
Obviously the Langevin analysis is not a full incorporation of
quantum fluctuations, which is a relatively unknown subject, but it
can provide phenomenologically trustworthy results. Definitely one
could try to incorporate quantum fluctuations differently, paying
the price of being model-dependent, but our above result seems
difficult to be changed.

\section{The double-screen Model}
\label{2screenm}

Single-screen entropic inflation is the simplest model inspired from
Verlinde's idea, and the scenario is interesting and totally
different from other inflation models. However, this scenario
presents difficulties in quantitatively describing the thermal
history of the universe due to the reason that the screen
temperature is always much lower than that of Cosmological Microwave
Background (CMB) and thus the universe would be unstable under
Hawking radiation. In order to alleviate such a problem, a
double-screen extension was proposed in
\cite{Cai:2010zw,Cai:2010kp}. In this model, apart from the usual,
outer holographic screen formed by the the Hubble horizon (or a
surface near it), an additional ``inner'' boundary has also been
introduced, which was considered as the Schwarzschild horizon of the
whole universe. At early time, both screens give rise to mutually
competing forces, which drive inflation, while at late times, when
the inner screen evaporates, the remaining outer screen drives the
universe acceleration.

The Schwarzschild radius $r_S$ is given by
\begin{eqnarray}
 r_S=2GM_{tot}=2G\int_{M_b}\rho dV=\frac{8\pi G\rho}{3\beta^3H^3},
\end{eqnarray}
where we have used that the volume of the universe is $V=4\pi
r_H^3/3$, with  $r_H=(\beta H)^{-1}$ the outer screen radius and
$\beta$ a dimensionless parameter quantifying the possible
divergence of the outer screen from the Hubble horizon. Its
corresponding temperature is given by
\begin{eqnarray}
 T_S=\frac{1}{8\pi GM_{tot}}=\frac{3\beta^3H^3}{32\pi^2G\rho},
\end{eqnarray}
and therefore its induced acceleration (with the simple entropy
form) will be $ a_S=2\pi T_S$, but with direction towards the inner
screen, opposite to the outer one. Therefore, in double-screen
entropic cosmology the total induced acceleration is
\begin{eqnarray}\label{accelaration}
 a_e=2\pi (T_H-T_S)
 =\beta H\bigg(1-\frac{3\beta^2H^2}{16\pi G\rho}\bigg),
\end{eqnarray}
with $T_H=\beta H/(2\pi)$, namely it incorporates a competition of
entropic effects from the outer and the inner screens. Taking also
the higher order corrections on the entropy expressions into
account, one extracts the modified Friedmann acceleration equation
in this scenario as \cite{Cai:2010zw,Cai:2010kp}:
\begin{eqnarray}
\label{friedmann3}
 \frac{\ddot a}{a} = -\frac{4\pi G}{3}(\rho+3p) + f(\rho, H)~,
\end{eqnarray}
with the form of surface function being
\begin{eqnarray}
 && f(\rho, H)
  \simeq \beta^2H^2\bigg(1-\frac{3\beta^2H^2}{16\pi G\rho}\bigg)\ \ \
  \ \ \ \ \ \ \ \ \ \ \  \ \ \ \ \ \   \ \ \ \ \ \  \ \ \
  \nonumber\\
  &&\ \ \  \ \  \ \  \
  \ +
\frac{g_HG\beta^4H^4}{4\pi}\left(1-\frac{27g_S\beta^6H^6}{
1024g_H\pi^3G^3\rho^3}\right)~,
\end{eqnarray}
 where $g_H$ and $g_S$ are
the corresponding dimensionless correction coefficients for each boundary.
Here we
have neglected higher order correction term that appeared in
\cite{Cai:2010zw,Cai:2010kp}. The cosmological system will close, as
usual, by the consideration of the evolution equation of the total
energy density $\rho$. In the case at hand, in which one may have
flow through the boundaries, the corresponding equation is modified
as \cite{Cai:2010zw,Cai:2010kp}
\begin{eqnarray}\label{HFRDBSCRIMPR2}
 \dot\rho+3H(\rho+P)=\Gamma~,
\end{eqnarray}
with the effective coupling term $\Gamma$ being
\begin{eqnarray}
 \Gamma = \frac{27\beta^6H^6}{1024\pi^3G^3\rho^3}\dot\rho
  +\frac{3\beta^2H\dot{H}}{4\pi G}
  \bigg(1-\frac{27\beta^4H^4}{256\pi^2G^2\rho^2}\bigg)~,
\end{eqnarray}
at classical level.

Focusing on the early-time universe evolution, and in particular on
the inflation realization, we assume that the universe is
radiation-dominated and thus the equation of state of the total
universe content is $P=\rho/3$. Solving the equations of motion
(\ref{friedmann3}) and (\ref{HFRDBSCRIMPR2}) up to leading order,
one can obtain the following approximate solution for the Hubble
parameter at early times  \cite{Cai:2010kp}
\begin{eqnarray}
H^2&\simeq&\frac{8\pi
G}{3}\bigg[\rho+\frac{8(g_H-4g_S)G^2}{69}\rho^2\bigg]
\nonumber\\
\label{HFRW}&=&\frac{8\pi
G}{3}\left(\rho+\frac{\rho^2}{\bar\rho}\right)~,
\end{eqnarray}
where $\bar\rho\equiv69/(8\bar{g}G^2)$, with $\bar{g}=g_H-4g_S$. An
interesting property of this scenario is that when $\bar{g}>0$, the
Hubble parameter is proportional to the energy density at high
energy scales. Therefore, in this case the $\rho^2$ term could make
the early time inflation much easier to be realized, comparing to
single-screen models.

\subsection{Inflation Eternality: Background Analysis}
\label{2screenback}

We now focus on the conditions for inflation eternality in
double-screen scenarios. Similarly to the previous section, we
consider that inflation will continue as long as condition
(\ref{slowroll}) holds, since it is model independent. In order to
simplify the expressions we restrict the analysis in the high energy
regime $\rho\gg\bar\rho$, where inflation is realized, since when
$\rho<\bar\rho$ inflation will always end \cite{Cai:2010kp}. In the
$\rho\gg\bar\rho$ case the second term in (\ref{HFRW}) dominates,
leading to a linear approximate relation between $\rho$ and $H$:
\be\label{approx}
\rho\simeq\sqrt{\frac{207}{64\pi\bar{g}}}\frac{H}{G^{3/2}}~. \ee
Substituting this into the second Friedmann equation
(\ref{friedmann3}) and considering also the radiation equation of
state $P=\rho/3$, we can eliminate $\rho$, resulting to an equation
depending only on $H$:
\begin{eqnarray}
\dot{H}=-\sqrt{\frac{23\pi}{\bar{g}G}}H+(\beta^{2}-1)H^{2}
-\frac{\beta^{4}}{2}\sqrt{\frac{\bar{g}G}{23\pi}}H^{3}\ \ \ \ \ \nonumber\\
\label{eq222}+\frac{g_{H}\beta^4}{
4\pi}GH^{4}-\frac{g_{S}\beta^{10}}{\pi^{3}}\frac{\bar{g}}{184}\sqrt{\frac{
\pi \bar{g}}{23}}G^{\frac{5}{2}}H^{7}.
\end{eqnarray}
Thus, condition (\ref{slowroll}) becomes:
\bea
\label{constraint}
-\sqrt{\frac{23\pi}{\bar{g}G}}H+\beta^{2}H^{2}&-&\frac{\beta^{4}}{2}\sqrt{
\frac{\bar{g}G}{23\pi}}H^{3}+\frac{g_{H}\beta^{4}}{4\pi}GH^{4}
\nonumber\\
&-&\frac{g_{S}\beta^{10}}{\pi^{3}}\frac{
\bar{g}}{184}\sqrt{\frac{\pi
\bar{g}}{23}}G^{\frac{5}{2}}H^{7}\gtrsim 0.\ \ \ \ \ \eea

In order to be more specific, as an example we consider the
parameter choice of \cite{Cai:2010zw}, that is $\beta=\sqrt{2}$
\footnote{This $\beta$ value ensures that at late times, when
the Friedmann equations become those of GR, that is $H^2=8\pi
G\rho/3$, the temperatures of the two screens will become equal, and thus
according to (\ref{accelaration}) the acceleration of the universe
caused by entropic force will vanish and inflation will exit
\cite{Cai:2010zw}. This
provides a exiting mechanism of inflation that is absent in single
screen model. }, $g_S=0$ and $\bar{g}=g_H=10^{16}$.
 Moreover, in the regime where
$\rho\gg\bar\rho$, the relation (\ref{approx}) leads to
$H\gg\sqrt{23\pi/(\bar{g}G)}$. Setting
$s=H\sqrt{\bar{g}G/(23\pi)}\gg1$, condition (\ref{constraint}) can
be simplified as: \be
H\left\{-1+23s\left[\left(s-\frac{1}{23}\right)^2+\frac{45}{529}\right]
\right\} \gtrsim 0.\ee Since $s\gg 1$, this condition is always
satisfied, that is, under the approximation $\rho\gg\bar\rho$, the
universe will not globally exit inflation. Note that this is
consistent with the result of \cite{Cai:2010kp}.

\subsection{Condition for Eternal Inflation: Thermal Fluctuation Analysis}

In the previous subsection we extracted the requirement for
inflation lasting at the background level, in the case of
double-screen scenarios. Here we desire to incorporate the thermal
fluctuations, and in particular to examine under what conditions the
inflation can be eternal. Similar to single-screen case, we will
estimate the change in energy density $\delta_t\rho$ due to thermal
fluctuations arising from normal radiation and from the holographic
screens, and we will compare it with the classical change
$\delta_c\rho$. The condition for eternal inflation will again be
$\delta_t\rho>\delta_c\rho$.

Concerning the classical change, things are different from
single-screen model due to the different background dynamics. In
particular, using (\ref{HFRDBSCRIMPR2}), (\ref{approx}) and
(\ref{eq222}), we obtain:
 \bea
\label{deltacrho2} \delta_c\rho&\approx&\frac{|\dot\rho|}{H}
=\frac{3[\frac{\beta^2\dot H}{4\pi G
}(1-\frac{27\beta^4H^4}{256\pi^2G^2\rho^2})-\frac{4\rho}{3}]}{1-\frac{
27\beta^6H^6}{1024\pi^3G^3\rho^3}}\nonumber\\
&\simeq&\frac{3H^2}{8\pi G
s^4}\left\{69s^3\left[\left(s-\frac{1}{23}\right)^2+\frac{22}{529}\right]
-3s^2+1\right\}.\ \ \ \ \  \eea However, concerning the thermal
fluctuations, the calculations are similar to the single-screen
scenario, since the mechanisms producing the fluctuations are
independent of the background. Therefore, in the next subsection we
study the conditions with both kinds of fluctuation productions.

\subsubsection{Normal radiation}

For standard radiation, the energy density writes as \be
\rho_{tr}=\frac{3g_r}{16\pi}T^4~,\ee where $g_r$ corresponds to the
term $64\pi\sigma{\cal G}(T)/3$ of the single-screen model.
Therefore, the heat capacity $C_V=g_rR_{tr}^3T^3$. Setting
$R_{tr}=c_s/H$ and $T=\beta H/2\pi$ we have: \be\label{deltaqrho21}
 \delta_t\rho|_{_{tr}}=\sqrt{\frac{g_r\beta^5}{(2\pi)^5c_s^3}}H^4~.\ee
Thus, comparing (\ref{deltacrho2}) and (\ref{deltaqrho21}) we deduce
that the condition for eternal inflation reads:
 {\small{\be
\label{condition1}
\sqrt{\frac{g_r\beta^5}{(2\pi)^5c_s^3}}H^4>\frac{3H^2}{8\pi G
s^4}\left\{69s^3\left[\left(s-\frac{1}{23}\right)^2+\frac{22}{529}\right]
-3s^2+1\right\}.\nonumber \ee}}
 This expression contains higher order term of $H$ and
$s=H\sqrt{\bar{g}G/(23\pi)}$, and thus since $s\gg 1$, it can be
simplified to give a constraint on $H(t)$ as:
\be\label{condition1approx} H>9\pi\sqrt{\frac{23\bar{g}
c_s^3}{2g_r\beta^5G}}~.\ee
 Since from this relation one can see that
$H^2>9\pi\sqrt{(23\bar{g} c_s^3)/(2g_r\beta^5G)}H\sim s/G\gg G^{-1}$
and thus $H\gg 1/\sqrt{G}$, eternal inflation can only be realized
at a very high energy regime.

\subsubsection{Holographic screen}

Concerning the holographic production of fluctuations, we can
consider that it can mainly be produced on the outer screen
\cite{Cai:2010kp}. Similar to the single-screen model, the total
energy of the screen is $\langle E\rangle\sim R_{hs}^2T/G$, the heat
capacity
 $C_V\sim c_vR_{hs}^2/G$, and thus
$\delta_t\rho|_{_{hs}}=T\sqrt{c_v/G}/R_{hs}^2$. Setting
$R_{hs}\simeq1/H$ and  $T=\beta H/2\pi$ (this $\beta$-modification
is the only difference between the two models) we finally obtain
\be\label{deltaqrho22}
\delta_t\rho|_{_{hs}}=\sqrt{\frac{c_v}{G}}\frac{\beta^3H^3}{2\pi}~.\ee

From (\ref{deltacrho2}) and (\ref{deltaqrho22}) we can see that the
eternality condition (\ref{eternal1}) turns out to be:
 {\small{
\be
\label{condition2}
\sqrt{\frac{c_v}{G}}\frac{\beta^3H^3}{2\pi}>\frac{3H^2}{8\pi G
s^4}\left\{69s^3\left[\left(s-\frac{1}{23}\right)^2+\frac{22}{529}\right]
-3s^2+1\right\},
\ee}}
 and in
the regime $s\gg 1$, it can be simplified as: \be
\label{condition2approx}
\left(\frac{4\beta^3\sqrt{c_vG}}{3}-69\sqrt{\frac{\bar{g}}{23\pi}}
\right)\sqrt { \frac {
\bar{g}}{23\pi}}H^2+6\sqrt{\frac{\bar{g}}{23\pi}}H-3>0~.\ee For
standard values of the parameters ($\bar{g}\gg G$,
$c_v,\beta\approx{\cal{O}}(1)$) the expression on the left hand side
does not accept non-trivial real roots, and thus condition
(\ref{condition2approx}) is never satisfied. Therefore, we conclude
that in the double-screen inflation model with the thermal
fluctuations from the holographic screen, the background result is
not changed.

Finally, if we desire to take into account simultaneously the
thermal fluctuations from standard radiation and from the
holographic screen, that is requiring
$\delta_t\rho|_{_{tr}}+\delta_t\rho|_{_{hs}}>\delta_c\rho$, and
under the approximation $s\gg1$, we result again to
(\ref{condition1approx})  since the holographic fluctuation
$\delta_t\rho|_{_{hs}}$ cannot have any contributions on eternal
inflation. Thus, this is the total condition for eternal inflation,
taking into account the thermal effect of fluctuations.

\subsection{Effects from Quantum Corrections: Langevin Analysis}

In the previous subsections we extracted the requirement for eternal
inflation in double-screen scenarios with both types of thermal
fluctuation productions. Similar to single-screen case, in the
present subsection we incorporate the micro-fluctuations, and we
examine whether they can prevent eternal inflation. In particular,
we perform a Langevin analysis in order to formulate the overall
cosmological evolution, that is the classical motion and the quantum
fluctuations, as a stochastic effect modeled through a random walk.

According to equation (\ref{eq222}) we can construct the
corresponding Langevin equation as:
 \be
\label{lagevin2} \dot{H}=C_1H+C_2H^{2}
+C_3H^{3}+C_4H^{4}+C_7H^{7}+q_dn(t)~, \ee
 where $q_d\sim\varepsilon G^{-3/4}$,
$n(t)$ satisfies (\ref{mean}), and the parameters $C_i(i=1,2,3,4,7)$
are:
\begin{eqnarray}
\label{c}
&&C_1=-\sqrt{\frac{23\pi}{\bar{g}G}}~,~C_2=\beta^{2}-1~,~C_3=-\frac{\beta^
{4}}{2}\sqrt{\frac{\bar{g}G}{23\pi}}~,
\nonumber\\
&&C_4=\frac{g_{H}\beta^{4}G}{4\pi} ~ ,
~C_7=-\frac{g_{S}\beta^{10}}{\pi^{3}}\frac{\bar{g}}{184}\sqrt{\frac{\pi
\bar{g}}{23}}G^\frac{5}{2}~,
\end{eqnarray} with dimensions of $1$, $0$, $-1$, $-2$ and $-5$,
respectively.

Now, imposing the expansion solution (\ref{expansion})  (with
$q_s\rightarrow q_d$), we obtain the separate equations \bea
\label{dhc2}
 \dot
H_c&=&C_1H_c+C_2H_c^2+C_3H_c^3+C_4H_c^4+C_7H_c^7,\\ \dot
H_1&=&C_1H_1+2C_2H_1H_c+3C_3H_1H_c^2+4C_4H_1H_c^3\nonumber\\
&\ &+7C_7H_1H_c^6+G^{-\frac{3}{4}}n, \\
\dot H_2&=&C_1H_2+C_2(H_1^2+2H_2H_c)
\nonumber\\
&\ &+3C_3H_c(H_1^2+H_2H_c) +2C_4H_c^2( 3H_1^2+2H_2
H_c)\ \ \nonumber\\
&\ & +7C_7H_c^5(3H_1^2+H_2H_c). \label{dhc2c} \eea These equations
cannot accept analytical solutions, however we can extract
approximate solutions in the limit $|t/{\cal T}|\ll 1$ with ${\cal
T}\gg H^{-1}$, that is performing our calculation within one
Hubble-interval ($t\sim H^{-1}$) \cite{Chen:2006hs}, and then we impose
the requirement
stochastic fluctuations to be able to stop a eternal inflation. The
explicit calculations are presented in the Appendix \ref{appendix2}.
We find that although it is very complicated to perform a general
analysis due to to high non-linearity of the equations, in the
specific parameter choice of \cite{Cai:2010kp} it is still hard for
the micro-fluctuations to have a significant effect on preventing
inflation from its eternality.

In summary, in this section we extracted the eternality conditions
for inflation in double-screen entropic scenarios, with thermal
fluctuations induced by normal radiation and holographic production.
As we saw, inflation is eternal in the high-energy regime.
Furthermore, we considered the quantum corrections from stochastic
effects by performing the Langevin analysis, which were shown to
hardly prevent the inflation eternality. On the other hand, in the
low-energy regime, inflation is globally exit \cite{Cai:2010kp}.

\section{Conclusions}
\label{Conclusions}

In this work we examined entropic cosmological scenarios, under the
light of inflation eternality. Going beyond Verlinde's basic idea,
which cannot describe inflation quantitatively, we considered both
the single-screen entropic inflation model \cite{Easson:2010xf}, in
which the holographic screen acts as an additional boundary term, as
well as its double-screen extension \cite{Cai:2010zw,Cai:2010kp},
where one adds a second, inner holographic screen. In particular,
after describing the inflation realization, we examined under what
conditions the inflation is eternal, taking into account both the
background evolution and mechanisms of thermal fluctuation
production. Furthermore, we incorporated quantum fluctuations
through a phenomenological, stochastic, Langevin analysis, and we
examined whether they can prevent the inflation from eternality.
Although the Langevin analysis is not a full incorporation of
quantum fluctuations, which is a relatively unknown subject, it can
provide phenomenologically trustworthy results.

After describing the inflation realization and the normal condition
for inflation to last at the background level, we discussed about
its eternality at the thermal fluctuations level. Firstly we defined
the general condition for eternal inflation as
$\delta_t\rho>\delta_c\rho$, which is very natural compared to the
well-known relation of scalar-field-driven inflation. In the case of
one-screen scenarios, we extracted the conditions for eternal
inflation with fluctuations from normal radiation and holographic
screen, (\ref{cond1}) and (\ref{cond2}) respectively, which should
be considered with (\ref{condsr}). These eternality conditions can
be easily fulfilled for general parameter choices. In the case of
double-screen model, we found that in high energy region when
$\rho\gg\bar\rho$ and inflation is always going on, the condition
for eternal inflation with radiation fluctuation remains
(\ref{condition1approx}), while for fluctuation produced by
holographic screen, eternal inflation is not affected. Finally,
after performing Langevin analysis, we found that it might be hard
in either of these two scenarios to prevent eternal inflation by
stochastic effects, unless severe fine-tuning the parameters. As a
side remark, however, we recognize that since the Langevin analysis
is still a random walk simulation instead of a complete calculation
of the quantum fluctuations, there could still be some ambiguity in
the validity of this result, which deserves more study \footnote{We
thank the referee for also pointing out this for us.}.

The above discussion indicates that from the cosmological point of
view one must be careful with entropic origin of gravity, since the
induced inflationary dynamics must have an exit, at least at some
parts of the universe, in order to be consistent with observations.
It seems that one-screen scenarios can easily lead to inflation
eternality, while double-screen considerations present a better
behavior. Finally, note that apart from these two models, in
\cite{Gu:2010wv} the authors discussed the case where the
holographic screen is open and the Brown-York surface stress tensor
is introduced, while in \cite{Nozari:2011et} noncommutative geometry
was used to describe the microstructure of the quantum spacetime of
entropic gravity. It would be interesting to investigate inflation
realization in such scenarios, and then focusing on the eternality
issues, examining whether the above behaviors can be improved.

However, the easy realization of eternal inflation in entropic
gravity could still be very interesting form the String Theory point
of view, since it may form the background for the ``Landscape'' of
string/M theory vacua, that is for the existence of a huge number of
possible (false) vacua
\cite{Kachru:2003aw,Bousso:2000xa,Susskind:2003kw,Douglas:2003um}.
Additionally, it may also be worth investigating the probability of
tunneling between different vacua, either Coleman-De Luccia type
\cite{Coleman:1980aw} or Hawking-Moss type \cite{Hawking:1981fz}. In
order to study these subjects, the full extension of our Langevin
analysis will become important. Such considerations may lead to new
perspectives in entropy gravity, and deserves further investigation.

\begin{acknowledgements}
The authors would like to thank Y. F. Cai, Yi Wang and especially Y.
S. Piao for useful discussions. They also thank the anonymous
referee for his helpful suggestions to the original version of the
paper. TQ is funded in part by the National Science Council of
R.O.C. under Grant No. NSC99-2112-M-033-005-MY3 and No.
NSC99-2811-M-033-008 and by the National Center for Theoretical
Sciences. ENS was supported in part by National Center of
Theoretical Science and  National Science Council
(NSC-98-2112-M-007-008-MY3) of R.O.C.
\end{acknowledgements}

\appendix

\section{Solution of the Langevin equations: The single-screen
scenario}\label{appendix1}

Since we are dealing with stochastic variables, we perform the
average of any physical quantity by defining the statistical
measure. In particular, we use the Fokker-Planck approach and define
the measure to be the physical volume of the Hubble patch, and thus
the average is defined as
\begin{equation}\label{average}
    \langle H(t)\rangle_p = \frac{\langle H(t)e^{3N(t)}\rangle}{\langle
e^{3N(t)} \rangle} \,,\quad N(t) = \int_0^t H(t')dt' \,.\nonumber
\end{equation}
Since the Hubble patch that is eternally inflating will have an
exponentially larger physical volume, taking the largest weight in
the average at late times, the physical volume can be a good measure
to characterize eternal inflation. Therefore, the average $\langle
H(t)\rangle_p $ could be significantly changed by stochastic
fluctuations if eternal inflation is realized. Furthermore, we shall
use the functional technique developed in \cite{Chen:2006hs} and
define a generating functional
\begin{equation}\label{generating fun}
    W_t[\mu] = \ln\langle e^{M_t[\mu]} \rangle \,, \quad M_t[\mu] =
\int_0^t \mu(t')H(t') dt' \,.\nonumber
\end{equation}
Thus, $\langle H(t)\rangle_p $ can be evaluated by functionally
differentiating $W_t[\mu]$ with respect to $\mu$ and setting
$\mu=3$, resulting to the following equations up to
$\mathcal{O}(q_s^2)$:
\begin{eqnarray}
  \langle H(t)\rangle_p  &=&  \frac{\delta W_t[\mu]}{\delta \mu(t)}
  \bigg|_{\mu(t)=3}\nonumber\\ &=&\langle H(t)\rangle + 3\int_0^t
\langle\langle H(t)H(t')\rangle\rangle dt' ,\ \ \ \ \ \label{HP}
\\
  \langle\langle H(t)H(t')\rangle\rangle &=& \langle H(t)H(t')\rangle
-\langle H(t)\rangle_p\langle H(t')\rangle_p \,.
\end{eqnarray}
After these definitions we can proceed to the solution of the
Langevin equations (\ref{dhc1})-(\ref{dhc13}).

In general, these equations cannot accept analytical solutions,
however we can obtain approximate solutions in the limit $|t/{\cal
T}|\ll 1$ (with ${\cal T}\gg H^{-1}$), that is performing our calculations
within one Hubble-interval ($t\sim H^{-1}$), where $t=0$ is the time where
inflation starts. In this case we can assume an
ansatz for the $H_c(t)$-solution, namely:
 \be \label{hc1}
 H_c(t)=H_{t0}+\tilde\Lambda_1 \left(\frac{t}{\cal T}\right)+\tilde\Lambda_2 \left(\frac{t}{\cal T}\right)^2...~,
\ee
 where
$H_{t0}\equiv H(t=0)$. Inserting this ansatz into equation (\ref{dhc1})
and by rescaling $\Lambda_1=\tilde\Lambda_1 {\cal T}^{-1}$,
$\Lambda_2=\tilde\Lambda_2 {\cal T}^{-2}$, we obtain: \be
 H_c(t)=H_{t0}+\Lambda_1 t+\Lambda_2 t^2...~,
\ee with
 \bea
 \label{h0}H_{t0}&=&\left(\frac{-15\pm\sqrt{57}}{28}\frac{D_1}{D_2}\right)^
{ \frac { 1 } { 2}}~,\nonumber\\
\label{lambda1}
\Lambda_1&=&D_1H_{t0}^2+D_2H_{t0}^4~,\nonumber\\
\label{lambda2} \Lambda_2&=&(D_1H_{t0}+2D_2H_{t0}^3)\Lambda_1~, \eea
where we remind that $D_1\equiv-\frac{2(1-C_H)}{1-2C_{\dot{H}}}<0$
and $D_2\equiv\frac{gG}{\pi(1-2C_{\dot{H}})}>0$. Note that
$\tilde\Lambda_{1,2}$ have dimensions of 1, while $\Lambda_1$ and
$\Lambda_2$ have dimensions of 2 and 3, consistently with the
dimensions of $D_1$ and $D_2$ as stressed before. One could also
check that this leads to $\Lambda_1<0$, which is asymptotically
quintessence-like inflation.

Thus, knowing the solution $H_c(t)$, we can acquire solutions of
$H_1(t)$ and $H_2(t)$ as
 \bea
\label{h11}
&&H_1(t)=h_{1i}E(t)\left\{1+\frac{1}{G^{3/4}h_{1i}}\int_0^t
n(t')E^{-1}(t')dt'\right\}\ \ \ \ \
\\
\label{h12} &&H_2(t)=E(t)\int_0^t \Delta(t)H_1^2E^{-1}(t')dt'~,
 \eea
where for convenience we have defined $E(t)\equiv
e^{\int_0^t(2D_1H_c+4D_2H_c^3)dt'}$ and $\Delta(t)\equiv
D_1+6D_2H_c^2$. Note that we have taken the initial condition for
the first-order correction of Hubble parameter to be $h_{1i}$, while
initial conditions for higher order corrections have been neglected.

Now, from (\ref{expansion}) we can straightforwardly write $ \langle
H(t)\rangle=H_c(t)+q_sG^{\frac{3}{4}}\langle
H_1(t)\rangle+q_s^2G^{\frac{3}{2}}\langle H_2(t)\rangle$, and
therefore (\ref{HP}) gives {\small{ \bea && \langle H(t)\rangle_p
=H_c(t)+q_sG^{\frac{3}{4}}\langle
H_1(t)\rangle+q_s^2G^{\frac{3}{2}}\langle
H_2(t)\rangle\nonumber\\
&&\ \ \ \ \ \ \ \ \ \  \ \
+3q_s^2G^{\frac{3}{2}}\int_0^t\Big[\langle
H_1(t)H_1(t')\rangle-\langle H_1(t)\rangle\langle
H_1(t')\rangle\Big]
dt'\nonumber\\
&&\ \ \ \ \ \ \ \ \ \ \
=H_{t0}+\Lambda_1t+q_sG^{\frac{3}{4}}h_{1i}E(t)+q_s^2G^{\frac{3}{2}}\left\{\langle
H_2(t)\rangle +\right.\nonumber\\
&&\left.\ \ \ \ \ \ \ \ \ \  \ \ 3\int_0^t\Big[\langle
H_1(t)H_1(t')\rangle-\langle H_1(t)\rangle\langle
H_1(t')\rangle\Big]dt'\right\},\ \ \ \ \ \ \ \label{express0} \eea}}
 where we have used that $\langle H_1(t)\rangle=h_{1i}E(t)$ and
that {\small{ \bea &&\langle H(t)\rangle\langle
H(t')\rangle=H_c(t)H_c(t')+  \ \ \  \ \ \ \ \  \ \ \ \ \ \ \ \ \ \ \
\ \  \
\ \ \ \ \ \ \ \  \ \ \ \ \ \ \ \ \  \ \ \ \ \ \ \ \ \ \nonumber\\
&&\ \ \ \   q_sG^{\frac{3}{4}}\left[H_c(t)\langle
H_1(t')\rangle +H_c(t')\langle H_1(t)\rangle\right]+\nonumber\\
&& \ \ \ \
 q_s^2G^{\frac{3}{2}}\left[\langle H_1(t)\rangle\langle H_1(t')\rangle
+H_c(t)\langle H_2(t')\rangle+H_c(t')\langle H_2(t)\rangle\right)]~.
\nonumber
 \eea}}
 The last terms on the right-hand side of (\ref{express0}) can be
expressed using {\small{
 \bea
&& \langle H(t)H(t')\rangle=H_c(t)H_c(t')+\ \ \  \ \ \ \ \  \ \ \ \
\ \ \ \ \ \ \ \ \  \
\ \ \ \ \ \ \ \  \ \ \ \ \ \ \ \ \  \ \ \ \ \ \ \ \ \ \nonumber\\
&&\ \ \ \   q_sG^{\frac{3}{4}}\left[H_c(t)\langle
H_1(t')\rangle+H_c(t')\langle H_1(t)\rangle\right]+\nonumber\\
&& \ \ \  \ q_s^2G^{\frac{3}{2}}\left[\langle H_1(t)H_ 1(t')\rangle
+H_c(t)\langle H_2(t')\rangle+H_c(t')\langle H_2(t)\rangle\right]~,
\nonumber
 \eea}}
and then, using relations (\ref{h11}), (\ref{h12}) and (\ref{mean}),
we obtain
\begin{widetext}
\bea
 &&\langle
H_1(t)H_1(t')\rangle=h_{1i}^2E(t)E(t')\left[1+\frac{1}{G^{3/2}h_{1i}^2}\int_0^{
min(t, t') }
E^{-2}(t_1)dt_1\right]\ \ \ \ \ \ \ \ \ \ \ \ \nonumber\\
&&\langle
H_2(t)\rangle=E(t)\int_0^t\Delta(t_1)E(t_1)\left[h_{1i}^2+G^{-\frac{3}{2}}\int_0^{t_1}E^{
-2} (t_2)dt_2\right]dt_1. \label{express1}
 \eea
\end{widetext}
Thus, inserting (\ref{express1}) into (\ref{express0}) and keeping
terms up to leading order in $E(t)$ and $H_c$ we acquire:
 \begin{eqnarray}
\langle
H(t)\rangle_p&=&H_{t0}+q_sG^{\frac{3}{4}}h_{1i}+\Lambda_1t\nonumber\\
&\ &+2q_sG^{\frac{3}{4}}h_{1i}H_{t0} (D_1+2D_2H_{t0}^2)t
\nonumber\\
&\ &+q_s^2G^{\frac{3}{2}}h_{1i}^2(D_1+6D_2H_{t0}^2)t~.
\label{HPfinal}\nonumber
 \end{eqnarray}
 This expression provides the
stochastic-fluctuation corrected Hubble-parameter evolution, namely
the first two terms on the right-hand side give the classical
result, while the last term provides the stochastic correction.
Therefore, requiring the stochastic-fluctuations to be able to
prevent eternal inflation, we need to impose {\small{
 \be
\label{condLang1}
2q_sG^{\frac{3}{4}}h_{1i}H_{t0}(D_1+2D_2H_{t0}^2)+q_s^2G^{\frac{3}{2}}h_{1i}^2(D_1+6D_2H_{t0}^2)
\lesssim \Lambda_1,
 \ee}} where $\Lambda_1$ is given in (\ref{lambda1}). This
expression provides the constraints on the initial condition of
stochastic fluctuations in leading order, namely $h_{1i}$, as:
 \be
 r_1^s\lesssim
h_{1i}\lesssim r_2^s~, \ee where {\small{ \be
r_{1,2}^s=-\frac{H_{t0}}{q_sG^{\frac{3}{4}}}\frac{(D_1+2D_2H_{t0}^2)\pm\sqrt{
2D_1^2+11D_1D_2H_{t0}^2+10D_2^2H_{t0}^4}}{D_1+6D_2H_{t0}^2}~.\nonumber
\ee}} Inserting in these expressions the definitions of $D_1$, $D_2$
and $q_s$ as functions of the parameters $C_H$ and $C_{\dot{H}}$
(see equation (\ref{lagevin1})), we find that in the required
intervals $C_H< 1$ and $0\leq C_{\dot H}\leq \frac{3}{4\pi}$
\cite{Easson:2010xf} there are no real solutions. Thus, condition
(\ref{condLang1}) cannot be satisfied and therefore stochastic
effects from quantum fluctuations cannot stop eternal inflation. We
mention here that performing the calculations within one Hubble
interval is a reasonable approximation, since the above result will
still be valid if one considers successively many Hubble intervals.

\section{Solution of the Langevin equations: The double-screen
scenario}\label{appendix2}

In this appendix we solve equations (\ref{dhc2})-(\ref{dhc2c}),
using also the introductory relations of appendix \ref{appendix1}.
In general, these equations cannot accept analytical solutions,
however we can acquire approximate solutions in the limit $|t/{\cal
T}|\ll 1$ (with ${\cal T}\gg H^{-1}$). Assuming an ansatz for the
$H_c(t)$-solution as
 \bea
H_c(t)&=&H_{t0}+\tilde\Sigma_1 \left(\frac{t}{\cal
T}\right)+\tilde\Sigma_2 \left(\frac{t}{\cal T}\right)^2...~,\nonumber\\
\label{hc2} &=&H_{t0}+\Sigma_1 t+\Sigma_2 t^2...~, \eea where in the
second step we use the similar rescaling as was done in Appendix
\ref{appendix1}, with $\tilde\Sigma_{1,2}$ of dimension 1 while
$\Sigma_1$ and $\Sigma_2$ of dimensions 2 and 3 respectively,
consistently with the dimensions of $C_i$'s stressed before.
Substituting it into (\ref{dhc2}) we obtain:
 \bea
\label{sigma1}
&&\Sigma_1=C_1H_{t0}+C_2H_{t0}^2+C_3H_{t0}^3+C_4H_{t0}^4+C_7H_{t0}^7\
\ \
 \ \ \nonumber\\
\label{sigma2} &&\Sigma_2=\frac{1}{2}\left(C_1+2C_2H_{t0}
+3C_3H_{t0}^2\right.\nonumber\\
&& \left.\ \ \ \ \ \ \  \ \ \ \ \ \ \  \ \ \ \ \ \ \ \ \ \ \
+4C_4H_{t0}^3+7C_7H_{t0}^6\right)\Sigma_1. \eea Thus, the solutions
for $H_1$ and $H_2$ read:
\begin{widetext}
 \bea
\label{h21}
&&H_1(t)=h_{1i}e^{\int_0^t\Xi_1(t')dt'}\left\{1+\frac{1}{G^{3/4}h_{1i}}\int_0^t
n(t')e^{-\int_0^{t'}\Xi_1(t'')dt''}dt'\right\}\ \ \ \ \ \\
\label{h22}
&&H_2(t)=e^{\int_0^t\Xi_2(t')dt'}\int_0^t\Pi(t')H_1^2e^{-\int_0^{t'}
\Xi_2(t'')dt''}dt'~,\ \ \ \ \ \eea
\end{widetext}
where \bea
&&\Xi_1(t)=C_1+2C_2H_c+3C_3H_c^2+4C_4H_c^3+7C_7H_c^6\nonumber\\
&&\Xi_2(t)=C_1+2C_2H_c+3C_3H_c^2+4C_4H_c^3+7C_7H_c^6\nonumber\\
&&\Pi(t)=C_2+3C_3H_c +6C_4H_c^2+21C_7H_c^5~. \nonumber \eea

Following the procedure of appendix \ref{appendix1} we finally
obtain:
\begin{widetext}
\bea \langle
H(t)\rangle_p&=&H_{t0}+\Sigma_1t+q_dG^{\frac{3}{4}}\langle
H_1(t)\rangle+q_d^2G^{\frac{3}{2}}\left\{\langle
H_2(t)\rangle+3\int\Big[\langle H_1(t)H_1(t')\rangle-\langle
H_1(t)\rangle\langle H_1(t')\rangle\Big]dt'\right\}
\nonumber\\
&=&H_{t0}+q_dG^{\frac{3}{4}}h_{2i}+\Sigma_1t+q_dG^{\frac{3}{4}}h_{2i}(C_1+2C_2H_{t0}+3C_3H_{t0}^2+4C_4H_
{t0}^3+7C_7H_{t0}^6)t\nonumber\\
&&+q_d^2G^{\frac{3}{2}}h_{2i}^2(C_2+3C_3H_{t0}+6C_4H_{t0}^2+21C_7H_{t0}^5)t~.\nonumber
\eea
\end{widetext}
Therefore, requiring the stochastic fluctuations to be able to
prevent the eternal inflation, we result to the constraint: {\small{
 \bea
&&\Sigma_1>q_dG^{\frac{3}{4}}h_{2i}(C_1+2C_2H_{t0}+3C_3H_{t0}^2+4C_4H_{t0}^3+7C_7H_{t0}
^6)\ \ \ \ \ \nonumber\\
&&\ \ \ \ \ \ \ \ \ \ \ \ \ \ \ \ \
+q_d^2G^{\frac{3}{2}}h_{2i}^2(C_2+3C_3H_{t0}+6C_4H_{t0}^2+21C_7H_{t0}^5)~.\nonumber
\eea}} Inserting the definition of $\Sigma$'s from  (\ref{sigma2})
and of $C_i$'s from (\ref{c}), we find that
 \be
 r_1^d\lesssim h_{2i}\lesssim
r_2^d, \label{lange2} \ee where
 \be
r_{1,2}^d=-\frac{X_0\pm\sqrt{X_0^2+4P_0\Sigma_1}}{2q_dG^{3/4}P_0},\nonumber
\ee
 with
\bea
 &&X_0=C_1+2C_2H_{t0}+3C_3H_{t0}^2+4C_4H_{t0}^3+7C_7H_{t0}^6,\nonumber\\
&&P_0=C_2+3C_3H_{t0}+6C_4H_{t0}^2+21C_7H_{t0}^5.\nonumber \eea
Relation (\ref{lange2}) imposes tight constraints on the model
parameters. For example in the particular parameter choice
$\beta=\sqrt{2}$, $g_s=0$, $g=g_H=10^{16}$ of \cite{Cai:2010zw} it
does not accept real solutions. Therefore, we conclude that on the
double-screen model at hand, the stochastic effects cannot induce an
exit from an eternal inflation caused by the background evolution
and thermal fluctuations, unless one tunes the model parameters
accordingly.

\end{document}